\documentclass{amia}

\usepackage{hyperref} 
\usepackage{listings}
\usepackage{graphicx}

\begin{document}
\title{Semantic Enrichment of Streaming Healthcare Data}
\author{Daniel Cotter, MS$^{1}$, V. K. Cody Bumgardner, PhD$^{2,3}$}

\institutes{
    $^1$Information Technology, University of Kentucky Medical Center, Lexington, Kentucky; $^2$Department of Pathology and Laboratory Medicine, Lexington, Kentucky; $^3$Department of Computer Science, Lexington, Kentucky\\
}

\maketitle

\noindent{\bf Abstract}

\textit{In the past decade, the healthcare industry has made significant advances in the digitization of patient information.
However, a lack of interoperability among healthcare systems still imposes a high cost to patients, hospitals, and insurers.
Currently, most systems pass messages using idiosyncratic messaging standards that require specialized knowledge to interpret. This increases the cost of systems integration and often puts more advanced uses of data out of reach.
In this project, we demonstrate how two open standards, FHIR and RDF, can be combined both to integrate data from disparate sources in real time and make that data queryable and susceptible to automated inference.
To validate the effectiveness of the semantic engine, we perform simulations of real-time data feeds and demonstrate how they can be combined and used by client-side applications with no knowledge of the underlying sources.}

\section*{Introduction}

\subsection*{Healthcare Information Technology and the Interoperability Problem}
Since the early 1970s, healthcare information technology has moved toward comprehensive electronic medical records (EMR) in which almost every aspect of the patient's healthcare has been digitized and retained indefinitely\cite{BensonGrieve2016}, which has vastly improved the efficiency with which patient information can be retained, communicated, and analyzed. At the same time, the healthcare industry has moved from a fee-for-service model to a value-based model, facilitated in part by the existence of such a record and in part by public policy, such as the Health Information Technology for Economic and Clinical Health (HITECH) Act of 2009 \cite{AdlerMilsteinJha2017}, which provided financial incentives for the "meaningful use" of electronic medical records.

The realization of a holistic medical record has been slowed by various obstacles, chief among them is the problem of interoperability between systems. The problem of interoperability arises almost as soon as a healthcare organization begins to choose a vendor for their electronic medical record, when they are faced with a choice between an architecture based on a single monolithic system or a so-called best-of-breed approach involving multiple discrete systems, each chosen for its superior performance in a narrow domain. The monolith claims to handle all aspects of healthcare information management; the best-of-breed approach entails a multiplicity of systems, each of which may be superior in its domain but which are not smoothly integrated.

A major difference between the two architectures is how they solve the problem of interoperability. In the case of the monolith, the problem is solved by the system vendor, at least in principle, but at the cost to the customer of a loss of choice. In the best-of-breed approach, the problem of interoperability is shifted onto the customer, who incurs an often hefty cost in the form of a more complex systems architecture and the resulting need for specialized hardware, software, and staff to maintain it.

In a best-of-breed approach, the need for instantaneous intersystems communication is typically handled via an Enterprise Service Bus (ESB)\cite{IBM_ESB}, which ensures real-time message delivery to subscribing systems. Additionally, if the data is to be analyzed in combination, rather than in isolation within the silo of a single system, it must be recombined and stored outside of these systems. This is typically done in an Enterprise Data Warehouse (EDW)\cite{KimballRoss2013} and requires further specialized hardware, software, and staff. However, most EDWs are based on a batch-loading system that runs during off-peak hours for the previous calendar day's business\cite{KimballRoss2013}; thus, while an EDW can be a powerful tool for retrospective analysis, it is unsuitable to real-time applications.

Interoperability is a serious challenge that modern healthcare systems must address in order to adequately serve their patients. In this paper we demonstrate a hitherto underused approach that combines the attractive aspects of both an enterprise service bus and an enterprise data warehouse to arrive at real-time analytics.

\section*{Background}
\subsection*{Health Level Seven Version 2 (HL7v2)}
HL7v2 is a healthcare messaging standard developed by the standards organization Health Level Seven International. It first emerged in 1988 and today is the most widely used such standard, having been adopted by over ninety-five percent of health systems in the United States and thirty-five countries worldwide \cite{HL7}. As such, it is something of a universal medium in the field of healthcare interoperability, yet it is terse and, without specialized training and access to the standard reference, cryptic.

Each HL7 message describes an event in a healthcare workflow and breaks down hierarchically into segments, fields, components, subcomponents, repeated components, and so on. There are well over one hundred types of messages and several times as many types of segments in HL7v2. The current version of the specification, for HL7 v2.8, is well over 2,500 pages long and contains nearly one million words. \cite{BensonGrieve2016} Partly as a consequence of this complexity, health interoperability has become a specialized field, replete with certifications and training and entire careers built on knowledge of HL7v2.  An example HL7 message describing the following information is shown in Figure \ref{fig:hl7v2.4}

\begin{itemize}
    \item The PID (Patient Identification) segment contains the demographic information of the patient. Eve E. Everywoman was born on 1962-03-20 and lives in Statesville OH. Her patient ID number (presumably assigned to her by the Good Health Hospital) is 555-44-4444.
    \item The OBR (Observation Request) segment identifies the observation as it was originally ordered: 15545 GLUCOSE. The observation was ordered by Particia Primary MD and performed by Howard Hippocrates MD.
    \item The OBX (Observation) segment contains the results of the observation: 182 mg/dl. 
\end{itemize}

\begin{figure}
  
\begin{lstlisting}[basicstyle=\footnotesize]
MSH|^~\&|GHH LAB|ELAB-3|GHH OE|BLDG4|200202150930||ORU^R01|CNTRL-3456|P|2.4<cr>
PID|||555-44-4444||EVERYWOMAN^EVE^E^^^^L|JONES|19620320|F|||153 FERNWOOD DR.^
^STATESVILLE^OH^35292||(206)3345232|(206)752-121||||AC555444444||67-A4335^OH^20030520<cr>
OBR|1|845439^GHH OE|1045813^GHH LAB|1554-5^GLUCOSE|||200202150730|||||||||
555-55-5555^PRIMARY^PATRICIA P^^^^MD^^|||||||||F||||||444-44-4444^HIPPOCRATES^HOWARD H^^^^MD<cr>
OBX|1|SN|1554-5^GLUCOSE^POST 12H CFST:MCNC:PT:SER/PLAS:QN||^182|mg/dl|70_105|H|||F<cr>
\end{lstlisting}
\caption{Example HL7 v2.4 Message}
\label{fig:hl7v2.4}  
\end{figure}

\subsection*{Health Level Seven Fast Healthcare Interoperability Resources (HL7 FHIR)}
FHIR \cite{FHIR} is a new open standard for healthcare data developed by the same company that developed HL7v2. However, whereas HL7v2 uses an idiosyncratic data exchange format, FHIR uses data exchange formats based on those already in wide use on the World-Wide Web such as Extensible Markup Language (XML) and JavaScript Object Notation (JSON) \cite{FHIR_DEV_OVERVIEW}, as well as the web's familiar transfer control protocols such as HyperText Transfer Protocol Secure (HTTPS) and Representational State Transfer (REST) \cite{FHIR_DEV_OVERVIEW} and system of contextual hyperlinks implemented with Uniform Resource Locators / Identifiers (URL/URI) \cite{FHIR_RESOURCES}. This design choice simplifies interoperability and discoverability and enables applications to be built rapidly on top of FHIR by the large number of engineers already familiar with web application design without a steep learning curve.

In contrast to HL7v2, which is based on events in a healthcare workflow such as admit, discharge, and transfer, FHIR is built on the notion of conceptual entities from the healthcare domain, such as Patient, Encounter, and Observation, i.e. \emph{resources}. Currently, FHIR encompasses 143 resources, each of which is described abstractly in the FHIR standard with the attributes Name, Flags, Cardinality, Type, and Description \& Constraints. \cite{FHIR_RESOURCES}. In a concrete implementation of FHIR, resources are serialized to one of the data exchange formats listed above.  An example of an FIHR XML message is shown in Figure \ref{fig:fihr}.

\begin{figure}
  
\begin{lstlisting}[basicstyle=\footnotesize]
<Bundle xmlns="http://hl7.org/fhir">
  <id value="b5be9983-c913-4d54-b2bb-8e848056662a"/>
  <type value="message"/>
  <entry>
    <resource>
      <MessageHeader>
        <id value="CNTRL-3456"/> <!-- ControlID of the v2 message -->
        <meta>
          <tag>
            <system value="urn:oid:2.16.840.1.113883.5.100"/>
            <code value="P"/>
            <display value="Production"/>
          </tag>
        </meta>
        <identifier value="CNTRL-3456"/>
        <timestamp value="2002-02-15T09:30:00-04:00"/>
        <event>
          <system value="http://hl7.org/fhir/message-type"/>
          <code value="observation-provide"/>
        </event>
        <source>
          <name value="GHH LAB"/>
          <endpoint value="urn:GHH-LAB"/>
        </source>
        <destination>
          <name value="GHH OE"/>
          <endpoint value="urn:GHH-OE"/>
        </destination>
        <data> <!--   The payload, the resource that this observation-provide concerns   -->
          <reference value="DiagnosticReport/1045813"/> <!-- Filler order number -->
        </data>
      </MessageHeader>
    </resource>
  </entry>
\end{lstlisting}
\caption{Example FHIR Bundle and Header Message}
\label{fig:fihr}  
\end{figure}

\subsection*{Semantic Web}
The term Semantic Web \cite{AllemangHendler2011} denotes an interconnected machine-readable network of information. In some ways it is analogous to the World-Wide Web, but with some crucial differences. The most important similarity is in the vision for the two technologies: Like the World-Wide Web, the Semantic Web was envisioned as a way for users from different institutions, countries, disciplines, etc. to exchange information openly and in doing so to add to the sum of human knowledge. The difference, however, is in the different emphases put on human readability versus machine readability: Whereas the World-Wide Web was intended to be visually rendered by one of any number of web browsers before being read by humans and therefore prioritizes fault tolerance and general compatibility over precision, the semantic web prioritizes precision and logical rigor in order for the information contained in it to be machine readable and used for logical inference.

The similarities continue in the technologies used to implement the two webs. Information in both the Semantic Web and the World-Wide Web is intended to be accessed using the familiar data exchange protocol Hypertext Transfer Protocol (HTTP) and addressed using Uniform Resource Identifiers (URI) for the Semantic Web and Uniform Resource Locations (URL) for the World-Wide Web that tell the agent/browser how to find linked information. Even the data exchange formats are remarkably similar: The World-Wide Web uses Hypertext Markup Language (HTML)\cite{HTML}, a tree-structured subset of Standard Generalized Markup Language (SGML)\cite{HTML_SPEC}, whereas the Semantic Web uses a variety of tree-structured formats such as XML, JSON, Terse RDF Triple Language (i.e. Turtle/TTL)\cite{TTL_SPEC}, etc.

The most significant difference between the World-Wide Web and the Semantic Web is in the type of information that they encode. The Semantic Web delivers a payload of simple logical statements known as triples, each consisting of a subject, predicate, and object, whereas the World-Wide Web delivers a series of directives to the web browser that govern the layout of the rendered page as well as the content of the page, in the form of text, images, videos, scripts, and so on. This difference in payloads corresponds to their different purposes -- the payload is delivered in the first case to an intelligent agent and in the second case to a web browser.

In more technical terms, the semantic web can be thought of as a distributed directed graph whose vertices are resources and whose edges are statements describing those resources. In its openness and decentralized nature, it bears some resemblance to the World Wide Web; however, whereas the World Wide Web consists of ad hoc, unsynchronized data presented in a variety of formats, the semantic web is a machine-readable body of information that can be synchronized while still coming from a variety of sources.

\subsection*{Resource Description Framework (RDF)}
RDF is the backbone of the semantic web\cite{AllemangHendler2011}. It is described as a framework, rather than a protocol or a standard, because it is an abstact model of information whose stated goal is "to define a mechanism for describing resources that makes no assumptions about a particular application domain, nor defines (a priori) the semantics of any application domain." \cite{RDF_SPEC} Its concrete realization is typically a serialization into one of several formats including XML, JSON, TTL, etc.

The basic unit of information in RDF is a \emph{statement} expressed as a logical triple; that is, a statement of the form \verb|<subject> <predicate> <object>|, in which the predicate expresses a relationship between the subject and the object: for instance, \verb|bloodPressure :value 120|. The subject must be a \emph{resource}, that is, an object consisting of one or more statements, and the object may be either a \emph{literal}, that is, a simple numeric or textual value, or another resource. The predicate describes some aspect or property of the subject. Because both the subject and the object can be resources, the object may also be described by statements in which it is the subject, leading to a complex graph structure.

A group of statements can be used to perform inference on their resources, thus creating new statements and enriching the semantic universe of the data set. For instance, the canonical syllogism "Socrates is a man; all men are mortal; therefore, Socrates is mortal" can be reproduced in the two statements \verb|Socrates :isA man| and \verb|man :is mortal|, resulting in a synthesized third statement: \verb|Socrates :is mortal|. RDF supports "inference, shared semantics across multiple standards and data formats, data integration, semantic data validation, compliance enforcement, SPARQL [SPARQL Protocol and RDF Query Language (SPARQL)] queries and other uses." \cite{FHIR_RDF}.

\subsection*{FHIR/RDF}
One of the several formats into which FHIR can be serialized is RDF. However, because RDF was designed as an abstract information model and FHIR was designed for operational use in a healthcare setting, there is the potential for a slight mismatch between the models. This comes up in two ways: One, RDF makes statements of fact, whereas FHIR makes records of events. The example given in the FHIR documentation is the difference between "patient x has viral pneumonia" (statement of fact) and "Dr. Jones diagnosed patient x with viral pneumonia" (record of event). Two, RDF is intended to have the property of \emph{monotonicity}, meaning that previous facts cannot be invalidated by new facts. The example given for this mismatch is "a modifier extension indicates that the surrounding element's meaning will likely be misunderstood if the modifier extension is not understood." The potential for serious error resulting from this mismatch is small, but it is worth bearing in mind when designing information systems.

\subsection*{SPARQL Protocol and RDF Query Language (SPARQL)}
RDF has an associated query language that can be used to search for matching statements, known as SPARQL. Although syntactically and semantically based on Structured Query Language (SQL), the information model over which it searches is RDF's directed graph of resources and statements, not the familiar relations stored in a relational database.

\begin{figure}[ht]
    \centering
    \includegraphics[width=1\textwidth]{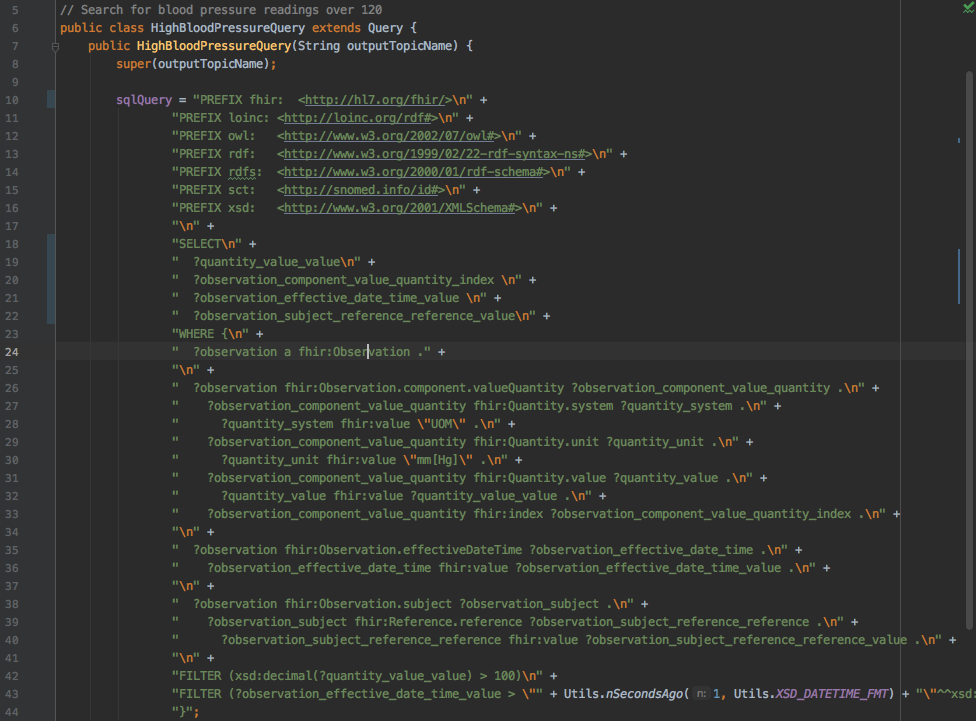}
    \caption{Example SPARQL Query}
    \label{fig:SPARQL_EXAMPLE}
\end{figure}

The syntax is beyond the scope of this paper, but in general SPARQL queries outline the shape of the graph they wish to find. For an example SPARQL query that searches for blood pressure readings over 120 b.p.m., see Figure \ref{fig:SPARQL_EXAMPLE}.

\section*{Method}
At a high level, the semantic enrichment engine is designed to take healthcare data in a variety of formats as input and store it in a triplestore database that users can query. In this way, the engine acts as both a collector, receiving messages from numerous sources, and a bus for delivering messages to multiple sources, as well as a real-time analytics platform. For example, a message from a vital signs monitor and from a registration system can be coalesced into a new stream containing blood pressure, temperature, and laboratory values for use in a machine learning model to predict sepsis.

\hbox{\hspace{2em}

\begin{picture}(400,150)
\put(-20,100){\framebox(150,20){pipe-delimited patients list}}
\put(-20,60){\framebox(150,20){JSON FHIR encounters}}
\put(-20,20){\framebox(150,20){HL7 observation messages}}
\put(140,40){\framebox(140,50){semantic enrichment engine}}
\put(290,100){\framebox(150,20){common data model}}
\put(290,60){\framebox(150,20){complex event processor}}
\put(290,20){\framebox(150,20){machine learning model}}
\put(130,100){\vector(1,-2){10}}
\put(130,70){\vector(1,0){10}}
\put(130,40){\vector(1,+2){10}}
\put(280,70){\vector(1,+4){10}}
\put(280,70){\vector(1,0){10}}
\put(280,70){\vector(1,-4){10}}
\end{picture}
}

 To support future large-scale operations, a multi-protocol message passing system was used for inter-module communication. This modular design also allows different components to be swapped out seamlessly, provided they continue to communicate via the expected interface.  Routines were developed to simulate input data based on the authors experience with real healthcare data. The reasons for this choice were twofold: One, healthcare data can be high in incidental complexity, requiring one-off code to handle unusual inputs, but not necessarily in such a way as to significantly alter the fundamental engineering choices in a semantic enrichment engine such as this one. Two, healthcare data is strictly regulated, and the process for obtaining access to healthcare data for research can be cumbersome and time-consuming. 

\begin{figure}[h]
    \centering
    \includegraphics[width=1\textwidth]{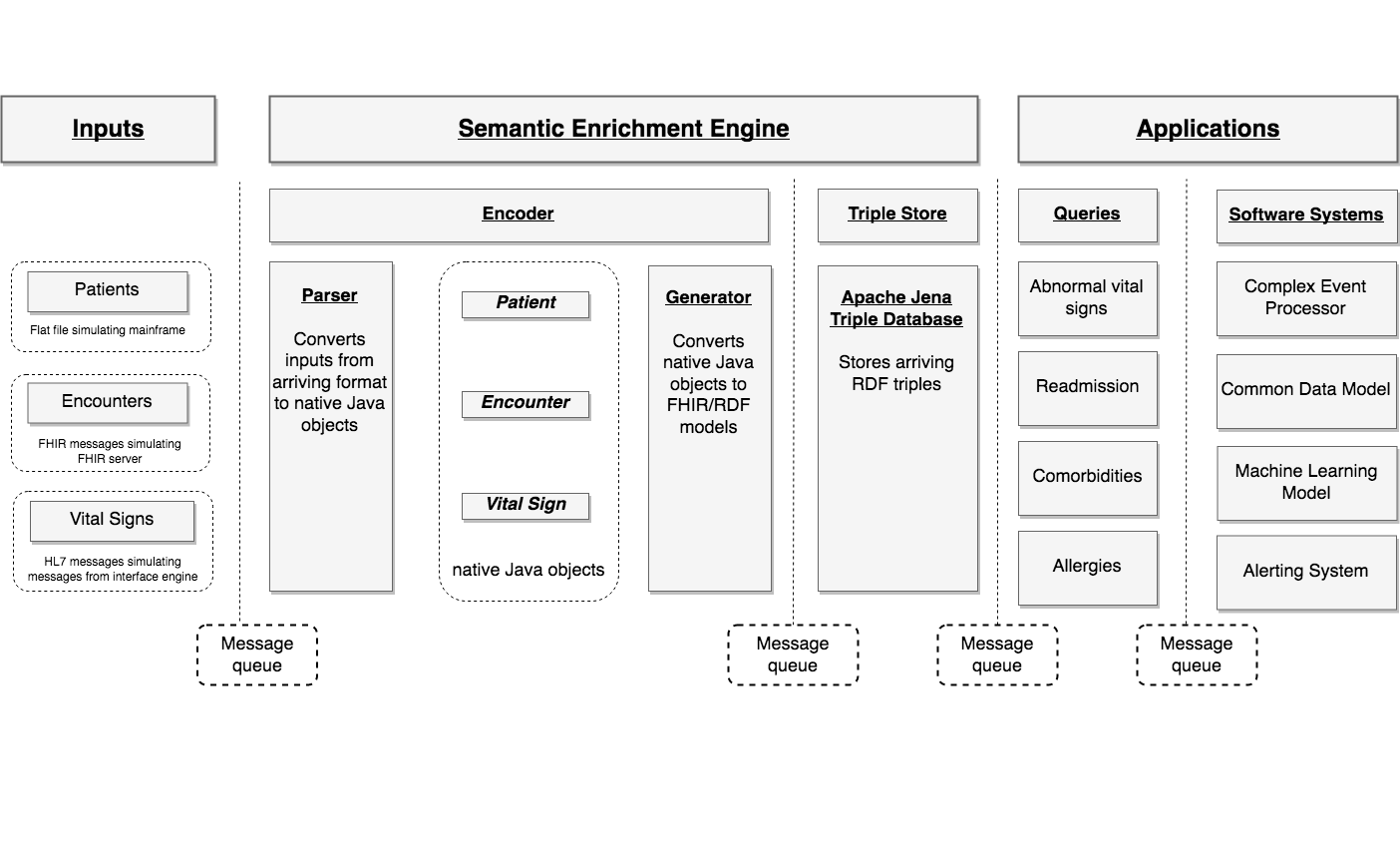}
    \caption{Semantic Enrichment Engine Architecture}
    \label{fig:high-level}
\end{figure}

A simplified set of input data, in a variety of different formats that occur frequently in a healthcare setting, was used for simulation. In a production setting, the Java module that generates simulation data would be replaced by either a data source that directly writes to the input message queue or a Java module that intercepts or extracts production data, transforms it as needed, and writes it to the input message queue.  A component-level view of the systems architecture is shown in Figure \ref{fig:high-level}

\subsection*{Class Hierarchy}
The project was written in Java, with each major component in its own package. There is a top-level class named ActiveMQEnabled that handles common tasks, such as connecting to the message broker, logging, event handling, and other such functionality. Each type of component in the pipeline - input, encoder, store, query, output, and application - is a subclass of ActiveMQEnabled as well as a superclass to specific types of those components. Most components are able both to send and receive messages, with certain exceptions: for example, inputs can only send and outputs can only receive. Stores can both receive and send, but in the concrete implementation in this project, the TDB store only receives (queries are better handled as timed polls, rather than being event-driven).

\subsection*{Inputs}
\begin{figure}[h]
    \centering
    \includegraphics[width=1\textwidth]{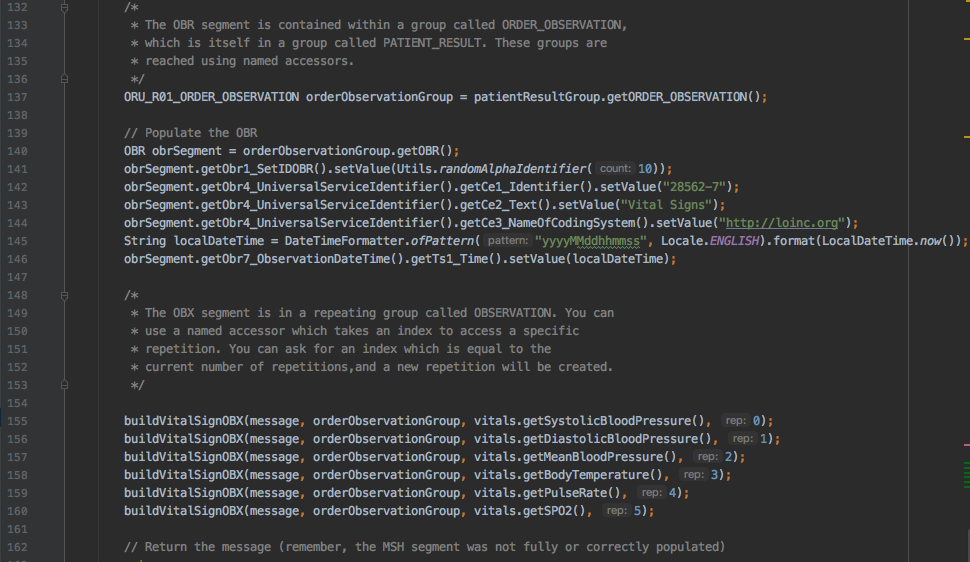}
    \caption{Java Simulated HL7 Message}
    \label{fig:input1}
\end{figure}

In the first stage of the module, simulated inputs represent a variety of healthcare entities and arrive in a variety of formats: patients in a pipe-delimited list, encounters as FHIR messages, and observations as HL7v2 messages. As discussed in the Background section, all of these are widely used input formats in modern health systems and realistically represent the heterogeneous message exchanges that are likely to occur in a real healthcare setting. Each input is configurable with regard to message timing and frequency, and the vitals signs can be made to simulate various conditions such as hypertension or hypothermia.  An example of a generate vital sign is shown in Figure \ref{fig:input1}

\subsection*{Encoder}
\begin{figure}[h]
    \centering
    \includegraphics[width=1\textwidth]{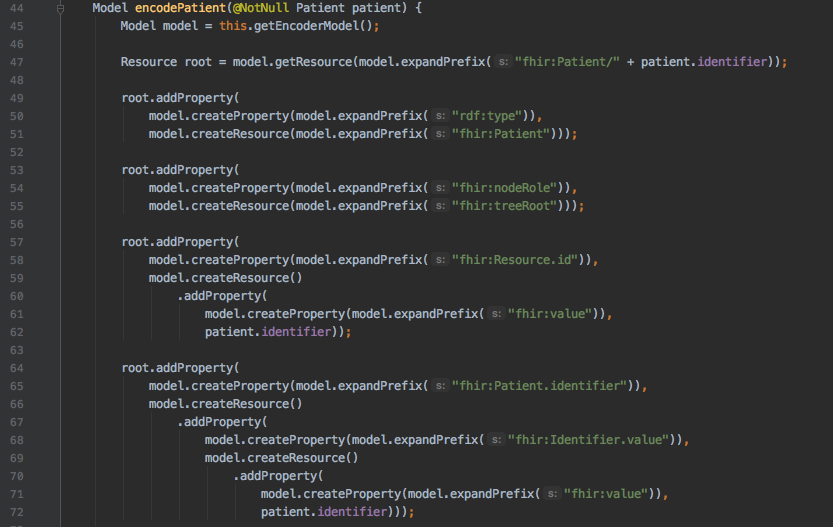}
    \caption{POJO to FIHR/RDF Encoder}
    \label{fig:encoder1}
\end{figure}

The encoder stage itself has two stages. In the first, input messages arriving at queues named according to the convention "INPUT.\emph{ENTITY}.\emph{FORMAT}" are retrieved, parsed, and transformed into internal representations of common domain objects, in this case Patient, Encounter, and Observation. In the second stage, these internal representations are transformed into internal representations of RDF graphs of FHIR resources and written out to the next message queue. By decoupling the parsing phase from the RDF-generating phase, the number of parsing and generating routines required for N sources and M resource types is reduced from N x M to N + M. This also allows parsing and generating jobs to be written in parallel and by different developers using the common internal representations as an intermediate layer. For instance, one developer could be writing the code to parse an HL7 ADT (admit/discharge/transfer) message while another developer was writing the code to turn this message into Patient, Encounter, and Observation resources. (Note that a single HL7 message can be used to create multiple FHIR resources \cite{FHIR_BLOG}).  An example of a POJO to FIHR/RDF message encoder class is shown in Figure \ref{fig:encoder1}

\subsection*{Store}
\begin{figure}[h]
    \centering
    \includegraphics[width=1\textwidth]{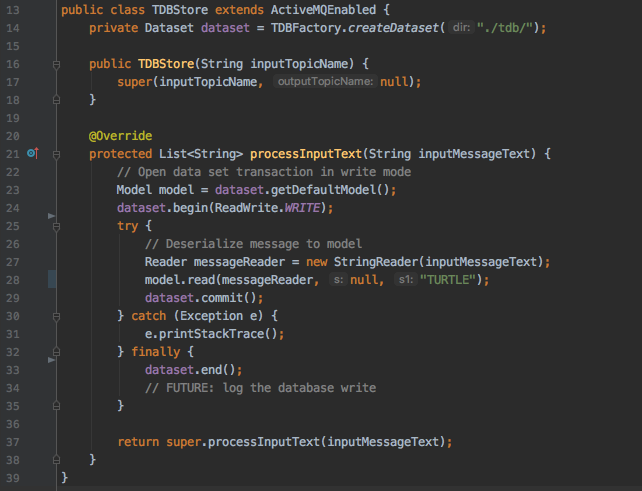}
    \caption{FIHR/RDF to TDB Storage Class}
    \label{fig:tdb}
\end{figure}

The store stage writes RDF-encoded statements to a triplestore database (TDB). For this implementation, the database was Apache Jena Triplestore Database (TDB) \cite{TDB}, which operates as a local on-disk database, although it could equally be a distributed in-memory cache or other implementation in production. It is at this point that the incoming messages are truly conformed to a universal model, as TDB does not record any information relating to encoding.  An example of a RDF to TDB (RDB Database) class is shown in Figure \ref{fig:tdb}

\subsection*{Query}
The query stage polls the triplestore database for RDF graphs matching specified criteria, for instance, low blood pressure combined with low body temperature and high pulse rate, indicating hypothermia, or patients with blood pressure readings over a certain threshold, indicating hypertension. It passes matching patients on to the output stage for data capture or immediate use in applications.

SPARQL queries against FHIR/RDF (see Figure \ref{fig:SPARQL_EXAMPLE}), can often be complex and verbose, simply because a high level of detail was required to represent healthcare data unambiguously in FHIR, and an equally high level of detail was required to extract it unambigously.

As a means of simplifying the work required to query the data, We considered a two-phase design in which the first layer would extract the relevant data from the TDB database in great detail before using RDF's \verb|CONSTRUCT| syntax to build a simplified representation of the data for use by the second layer. This idea has potential, but after a few tries at writing the code to implement it, there was too much loss of detail for it to be worth pursuing in this iteration. In the end, the default option of writing a detailed, if verbose, RDF query once was deemed a better option than the added complexity and potential loss of fidelity of the two-layer approach.

\subsection*{Output}
In the output stage, the results of the queries in the previous stage are written out to an output destination such as a text file or a screen. This differs from the Application stage in that the input was intended to be written immediately to an output sink such as a file or screen on the local computer. Its use in this project was limited to debugging.

\subsection*{Application}
\begin{figure}[h]
    \centering
    \includegraphics[width=1\textwidth]{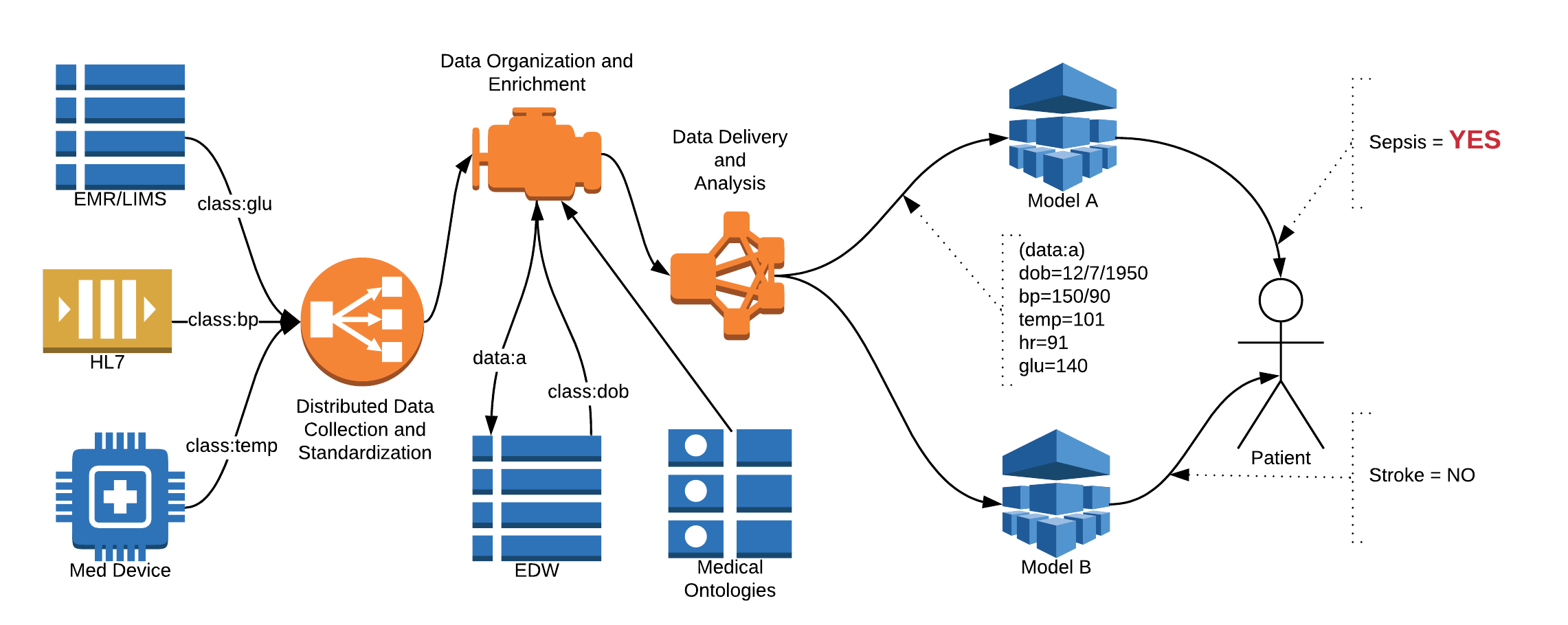}
    \caption{Semantic Engine Use Clinical Workflow}
    \label{fig:high-level-app}
\end{figure}

In the application stage, a variety of applications (complex event processors, common data models, machine learning models, etc.) receive the outputs of the queries from the prior stages and use them as inputs to particular applications.  A high-level view of how the semantic encoder might be used in clinical workflow is shown in Figure \ref{fig:high-level-app} 

Several applications presented themselves as potentially benefiting from a semantic enrichment engine such as this one. One such application was complex event processing (CEP), in which streams of data are analyzed in search of \emph{events} in real time\cite{Luckham2012}. From simple events more complex events can be derived, so that a number of individually innocuous events may add up to either an opportunity or a threat event. In a healthcare setting, this could mean monitoring patient vital signs and flagging them as high, low, or normal, then analyzing the combination of vital signs for a condition or set of conditions. Additionally, a patient's individual health conditions, such as comorbidities, recent procedures, and so on could be used to inform the meaning of the instantaneous vital signs as they are received. Using data from the TDB store, I was able to write several queries in Esper, a well-known complex event processing engine\cite{ESPER}, to detect conditions that were initially simulated by the vital signs input, such as hypothermia or hypertension. To some extent, the RDF queries used to feed Esper overlapped with the capabilities of Esper itself, although Esper's query language EPL is much more versatile than SPARQL for event processing.

Another such project was the Observational Medical Outcomes Partnership (OMOP) Common Data Model (CDM)\cite{OHDSI_OMOP_CDM}. This is an analytical database intended to collate data from multiple partner data sources and conform it to a common representation, using standardized vocabularies such as LOINC\cite{LOINC} and SNOMED-CT\cite{SNOMED} in order to facilitate collaborative research. Using data queried from the TDB store, I was able to build several data-loading jobs to populate an OMOP-CDM database. This application takes advantage of the semantic enrichment engine's ability to conform data from disparate sources, since by the application stage all the data has been conformed to FHIR/RDF and is ready to be loaded to the OMOP database with only one transformation (from FHIR/RDF to OMOP schemas).

\subsection*{Validation}
Health Level Seven International (HL7) provides a FHIR validator, which was useful for ensuring that the FHIR generated by the encoder was correctly formed. ShEx (Shape Expressions) \cite{SHEX} language is a language used for describing the expected shape of RDF and testing it for conformity to that shape. Its syntax is similar to Turtle and SPARQL, while its semantics resemble those of regular expression languages such as RelaxNG \cite{RELAX_NG}. I were limited in our ability to validate FHIR conformance due to limitations of the FHIR validation tool (vague error messages, program crashes, etc.)

\subsection*{Challenges}
Our needs are twofold and, at first, apparently contradictory. The first was to store data from disparate sources so that the sources could be joined up and benefit from synergies among the different semantic components embedded in the data. The second was to answer queries about the data over a finite time range. The challenge is that the mechanism that was to trigger the execution of a query, the receipt of a message from the store, happened with such frequency that the query engine quickly became overloaded and unable to respond in a timely fashion to new requests. This necessitated a redesign of parts of the encoder module and the query engine, such that each resource was timestamped when it was encoded and each query specified a time range within which to return results. Prior to this redesign, the query engine was querying the triple store each time a message arrived without specifying a time bound, resulting in a constantly increasing number of results that eventually would overmatch the system's capabilities.

Another challenge was that RDF does not easily support streams\cite{W3}. With each query, all matching results are returned, not only the new results since the last query. This means the result size of the query increases monotonically until the system is overwhelmed. To design around this, we timestamped each entity as it arrived and used this timestamp as a filter in the subsequent queries. This worked well and is not unlike what CEP systems do natively.


\section*{Conclusion}
The semantic enrichment engine designed described in this paper has broad applicability in healthcare operations and research.  The data exchange standards, protocols, databases, query languages, and so forth used to implement this system are freely available. This system has characteristics of both an enterprise service bus and an enterprise data warehouse, but augments the analytical capability of the former and  addresses the high latency of the former. We expect the system can be used to inform artificial intelligence for inference, populate structured databases with enriched data streams, and derive new data for use in machine learning training. 

\makeatletter
\renewcommand{\@biblabel}[1]{\hfill #1.}
\makeatother

\bibliographystyle{unsrt}
\bibliography{daniel.bib}

\end{document}